# Initial validation of a virtual-reality learning environment for prostate biopsies: realism matters!

Gaelle Fiard[1,2], Sonia-Yuki Selmi[2], Emmanuel Promayon[2], Lucile Vadcard[3], Jean-Luc Descotes[1,2] and Jocelyne Troccaz[2,1]

[1] Department of Urology, Grenoble University Hospital, Grenoble, France
[2] TIMC-IMAG Laboratory, CNRS UMR 5525, UJF Grenoble, France
[3] Laboratory for Educational Science, UPMF, Grenoble, France

Word count: 2139

Abstract: 273

**Address for correspondence:**

Dr Gaelle Fiard

Department of Urology

Grenoble University Hospital

38043 Grenoble cedex 9, France

GFiard@chu-grenoble.fr

**Funding sources:**
This work was supported by French state funds managed by the ANR within the PROSBOT project under reference ANR-11-TECS-0017 and the Investissements d'Avenir programme (Labex CAMI) under reference ANR-11-LABX-0004, the Association Française d'Urologie - AstraZeneca (grant G.Fiard) and by INSERM CHRT (grant J.Troccaz).

# Abstract

**Introduction-objectives:** A virtual-reality learning environment dedicated to prostate biopsies was designed to overcome the limitations of current classical teaching methods. The aim of this study was to validate reliability, face, content and construct of the simulator.

**Materials and methods:** The simulator is composed of a) a laptop computer, b) a haptic device with a stylus that mimics the ultrasound probe, c) a clinical case database including three dimensional (3D) ultrasound volumes and patient data and d) a learning environment with a set of progressive exercises that comprises a randomized 12-core biopsy procedure. Both a visual feedback (3D biopsy mapping) and numerical feedback (score) are given to the user. The simulator evaluation was conducted in an academic urology department on 7 experts and 14 novices who each performed a virtual biopsy procedure and completed a face and content validity questionnaire.

**Results:** The overall realism of the biopsy procedure was rated at a median of 9/10 by non-experts (7.1-9.8). Experts rated the usefulness of the simulator for the initial training of urologists at 8.2/10 (7.9-8.3), but reported the range of motion and force feedback as significantly less realistic than novices (p=0.01 and 0.03 respectively). Pearson's r correlation coefficient between correctly placed biopsies on the right and left side of the prostate for each user was 0.79 (p<0.001). The 7 experts had a median score of 64% (59-73), and the 14 novices a median score of 52% (43-67), without reaching statistical significance (p=0,19).

**Conclusion:** The newly designed virtual reality learning environment proved its versatility and its reliability, face and content were validated. Demonstrating the construct validity will require Improvements to the realism and scoring system used.

## Introduction

Current teaching methods for prostate biopsies, based on an apprenticeship without feedback on biopsy distribution nor overall performance have shown their limitations. Previous publications reported discrepancies between the assumed location of cancer foci based on biopsies and their actual location on final pathology (1). Adding a visual feedback proved to improve the biopsy distribution within the gland, even for an experienced operator (2).

The emergence of targeted biopsies and focal therapy created a need for a better sampling of the prostate to avoid leaving cancer foci untreated, and new training needs to target predefined areas of the prostate, using mental reconstruction or Magnetic Resonance Imaging (MRI)-transrectal ultrasound (TRUS) fusion systems (3). Simulation-based training could therefore find a place in the initial training, but also in the performance assessment of these new techniques.

Various simulators already exist in the field of urology, although most are dedicated to laparoscopy and, as stated by a recent report by the French Health Authorities (HAS), are also largely underused (4)(5). A simulator for prostate biopsies was already described, focused more on the gesture itself than the entire biopsy procedure, and with a limited teaching environment (6).

We designed a simulator dedicated to prostate biopsies, allowing visual feedback and performance assessment, enhanced by a complete learning environment and connected to a clinical case database in order to cover the main situations encountered in clinical practice (7)(8). The aim of this study was to validate the reliability (reproducibility and precision), face (the simulator represents what it is supposed to represent), content (the simulator teaches what it is supposed to teach) and construct (the simulator is able to distinguish between experienced and inexperienced users) of the simulator.

## Materials and methods

**1. Simulator design**

1.1 <u>Simulator architecture</u>

The simulator is composed of a laptop computer connected to a haptic device (Phantom Omni, Sensable) with a stylus held as the ultrasound probe (Figure 1). Manipulation of the stylus allows the user to explore the ultrasound volume and display the corresponding two-dimensional (2D) ultrasound image. Three dimensional (3D) TRUS volumes with corresponding prostate mesh (prostate segmentation) acquired with an end-firing 3D endorectal probe (SonoAce X8, Samsung-Medison) during actual biopsy procedures with the Urostation® (Koelis, La Tronche, France) were collected and entered after anonymization into a clinical case database. Pertinent clinical information (age, prostate volume, digital rectal examination, Prostatic Specific Antigen level) and prostate MRI were also collected for each case, when available.

**Figure 1.** Simulator design.

## 1.2 Learning environment

The various exercises and functionalities of the simulator were developed based on a didactical study performed at the beginning of the project. The learning outcomes expected include technical skills (ability to distribute randomized biopsies to obtain an optimal coverage of the gland, and to target suspicious areas), theoretical knowledge and decision-making ability (cancer risk estimation based on clinical data and PSA level). The exercises therefore include the practice of every aspect of the biopsy procedure: TRUS image reading, prostate volume measurement, prostate cancer risk estimation, sector or area targeting, and MRI-TRUS mental fusion. Each exercise can be performed with or without assistance. Both visual feedback (3D visualization and assistance) and numerical feedback (biopsy score) are provided. Assistance consists in displaying a 3D representation of the prostate surface in which the 2D ultrasound plane can be visualized in real time. Additionally, the user can choose to display the needle trajectory and the already performed biopsies (Figure 2). All these exercises aim at helping trainees to build a 3D mental representation of the prostate based on 2D images, and to improve their hand-eye coordination. Each exercise can be performed independently, or can be combined into a complete learning pathway. Learning pathways are adjusted to each user's specific training needs depending on their initial evaluations.

**Figure 2.** Biopsy assistance: 2D ultrasound plane (a), needle trajectory (b), previous biopsies (c).

1.3 Biopsy procedure

Virtual biopsies can be performed and located within the prostate volume. A biopsy procedure simulation exercise replicates the pertinent conditions of a real 12-core biopsy procedure. Users have to perform a preoperative checklist, 12 virtual biopsies using the left hand (or right hand for left handed users) for biopsy firing without immediate feedback, and have to specify the location of each core as if they were preparing it for the pathologist. The target of each of the twelve cores is one of the twelve sectors as represented in Figure 3. At the end of the procedure a visual feedback and a numerical feedback are shown. The visual feedback consists in the representation of each biopsy core within the prostate volume, and the numerical feedback in a percentage score (see section 1.4), as shown in Figure 3.

1.4 Scoring system

To provide the user with a numerical feedback on his/her performance, a simple scoring system has been developed. The prostate volume is divided into twelve parts (sectors) of equal volume, as shown in Figure 3. Each biopsy is awarded 4 points when the targeted sector is actually reached, and only 1 point when it is inside the prostate but outside the targeted sector. The scores of the 12 biopsies are added and the result converted to a percentage score. This score is displayed to the user along with the visual feedback (Figure 3a bottom left).

**Figure 3.** Visual feedback (biopsy mapping) and numerical feedback (score) (a). Division of the prostate volume into twelve sectors of equal volume (b).

## 2. Evaluation

### 1.1 Evaluation settings

The evaluation was conducted in an academic urology department among 14 novices (medical students and non-urologist residents) and 7 experts (senior urologists). All experts had performed more than 50 biopsy procedures. The simulator and the prostate biopsy procedure were first presented during a briefing session. A login was created for each user in order to individually record the duration of the biopsy procedure, score, and position of each biopsy for future reviewing. An initiation phase with standardized steps allowed the users to apprehend the manipulation of the stylus that mimics the ultrasound probe. They were then asked to perform a virtual standard randomized 12-core biopsy procedure, without assistance or immediate feedback on the location of the biopsies. Their overall performance and score were recorded. Each user was finally asked to fill an anonymous questionnaire.

### 1.2 Face and content validity

The face validity was evaluated through the aforementioned anonymous questionnaire. The overall realism of the virtual biopsy procedure, and the realism of the US image, of the range of motion of the US probe and of the force feedback were assessed using a Visual Analogue Scale (VAS).

The content validity was evaluated by the 7 experts who were also asked, after completing their biopsy procedure, to go through the various exercises offered by the simulator and evaluate their interests in the initial training of urology residents. This was also done using a VAS. Median values and interquartile range were computed. The detailed questionnaire used is available in **Appendix 1.**

### 1.3 Intrinsic reliability

To assess the intrinsic reliability of the simulator, the split-halves technique was used, comparing performance of subjects on one side of a test with performance on the other side (9). The number of correctly placed biopsies (biopsies reaching the targeted sector) on the right side of the prostate was therefore compared to the number on the left side for each user using Pearson's r correlation. Mean values and 95% Confidence Interval (95%CI) were computed.

### 1.3 Construct validity

To evaluate the ability of the simulator to discriminate between experts and novices, we recorded the scores of the 7 experts and 14 novices were recorded and compared using the Mann-Whitney U test.

### 1.4 Statistical analysis

Statistical analysis was performed using *R* statistical software (version 2.13.1 for Mac; The R Foundation for Statistical Computing, http://www.R-project.org), with statistical significance defined at $p<0.05$.

## Results

**1. Face and content validity**

The overall realism of the biopsy procedure was rated at a median of 9 for an interquartile range of (7.1-9.7), the quality of the graphical user interface at 9.1 (7.8-10), the realism of the US image at 9.1 (7.1-10), the realism of the range of motion of the ultrasound probe (stylus) at 7.5 (4.9-9.1), and the realism of the force feedback at 5.6 (4.9-8.8). The detailed results of the ratings attributed by novices and experts are shown in Figure 4. Experts reported the range of motion and force feedback as significantly less realistic than novices (p=0.01 and p=0.03, respectively).

The 7 experts rated the usefulness in the initial training of urology residents at 8.2 (7.9-9.6), as represented in Figure 4.

**Figure 4.** Face and content validity questionnaire: evaluation of the graphical user interface (a), the realism of the ultrasound image (b), range of motion (c), force feedback (d), biopsy procedure (e) and the usefulness of the simulator for initial training (f).

### 2. Intrinsic reliability

The mean number of biopsies correctly reaching the targeted sector on the right side was 2.8 (95%CI 2.2-3.4), and 2.7 on the left side (95%CI 2.0-3.5). The Pearson's r correlation coefficient for each user between the results on the right and left side was r=0.79 (95%CI 0.55-0.91), with p<0.001. Considering the results of the experts only, the number of biopsies correctly placed on the right side was 3.1 (95%CI 2.1-4.1), and 3 (95%CI 1.5-4.5) on the left side. Pearson's r correlation for experts only was r=0.89 (95%CI 0.44-0.98), with p=0.006.

### 3. Construct validity

The median score for the 21 users was 60% (43-68). The 7 experts had a median score of 64% (59-73), and the 14 novices a median score of 52% (43-67), difference in which did not prove to be statistically significant (p=0,19). Results are displayed in Figure 5.

**Figure 5.** Comparison of the scores obtained by experts and novices.

## Discussion

The results of this initial evaluation allowed us to validate the face and content of our simulator, with a median rating of 9/10 by nonexperts for the realism of the biopsy procedure, and a median rating of 8.2/10 by experts for the educational interest of the simulator. The split-halves technique showed a correlation coefficient between the right and left side of the prostate for each user of 0.79. This correlation coefficient represents the proportion of variability in scores attributable to true differences between users. This result is very close to 0.8, the value considered as satisfactory to avoid misclassification and to validate the intrinsic reliability of the simulator (9).

We failed to prove the construct validity despite a 12% difference in the median scores of novices and experts. This can be partly explained by the small size of the panel and the important variability of the scores in each subgroup. Taking into account this variability, we calculated a posteriori that 44 users would have been required in each group to statistically prove a 12% difference with 80% power.

Another explanation can be found in the very simple scoring system that was used, which was probably insufficient to properly discriminate between the 2 groups. Defining the criteria of a "good" procedure is central to the design of a simulator and in itself a very interesting contribution to medical practice assessment. The main advantage of the simple score we used is that it is easily understandable by the users. We are now working on the creation of a more complex score based not only on the reached sectors, but also on other factors. They include the resulting length of the needle inside each sector, the distance from the targeted sector when it was not reached, and the total duration of the procedure. This new score should allow better discrimination between novices and experts.

This evaluation also showed that the simulator could improve its realism. In its current version, the user is required to manipulate a small size round-shaped symmetrical stylus in place of an ultrasound probe. Although a recent review comparing low- and high-fidelity simulators showed only a small learning improvement when realism is increased, the interaction with the stylus was particularly confusing for the experts (10). They had trouble maintaining a steady axial plane during their virtual biopsy procedure. This was not the case for the novices who were not used to use a real ultrasound probe. Also, the 2D images reconstructed from the exploration of the 3D ultrasound volumes were not always strictly similar to the 2D images obtained with a classical 2D US probe. This can probably explain the lower ratings of the experts compared to the novices regarding certain aspects of the realism (probe range of motion, force feedback). This can also partly explain the smaller than expected difference of the scores of the experts compared to novices. In the planned new version of the simulator, a mockup of a real US probe, obtained using a 3D printer, will replace the stylus.

This is to our knowledge the second simulator for prostate biopsies described in the literature, but the first one with such an educational environment and performance evaluation (6). This initial study aimed at validating the simulator only in its ability to simulate a 12-core randomized biopsy procedure. But it also offers other functionalities, notably a learning pathway enabled to propose exercises that meet the user specific training needs. This learning pathway, as well as the ability of the trainee to reproduce the same performance on an actual patient (predictive validity), will also have to be further validated.

## Conclusion

Validating a simulator is a complex task that should take into account not only the intrinsic properties of the simulator, but also its realism, the scoring system used and the conditions of its validation. The development of such a simulator requires the characterization of the quality of a gesture and the description of the required skills and simulated procedures. We designed a learning environment for prostate biopsies, which proved its reliability, and we validated its face and content. Improvements to the realism and scoring system used are necessary to validate the construct. Further studies will also be required to evaluate the learning pathway and predictive validity of this simulator.

## Acknowledgments

We want to thank Dr Pierre Mozer, from La Pitié Salpétrière Hospital, Paris, for his strong involvement in the early versions of Biopsym and Koelis for giving us access to internal data structures of the Urostation®.

## Author Disclosure Statement

No competing financial interests exist.

Figure 1.

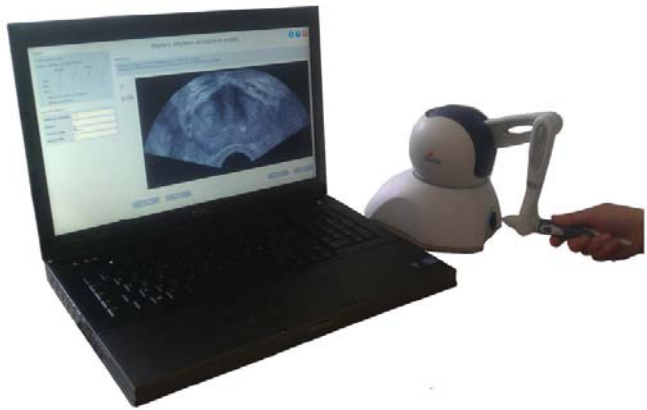

Figure 2.

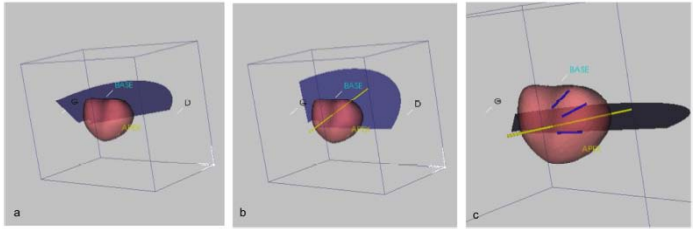

Figure 3.

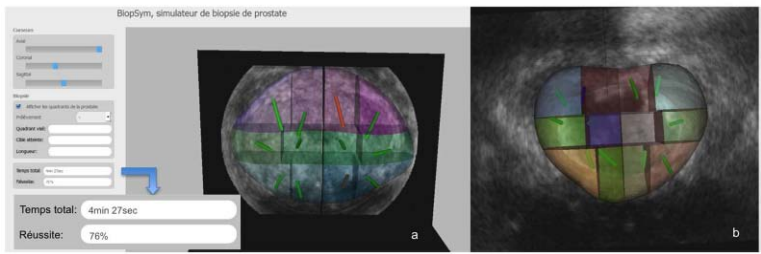

Figure 4.

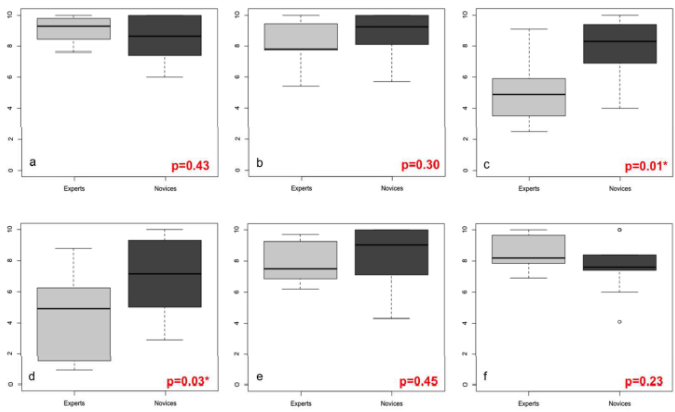

**Figure 5.**

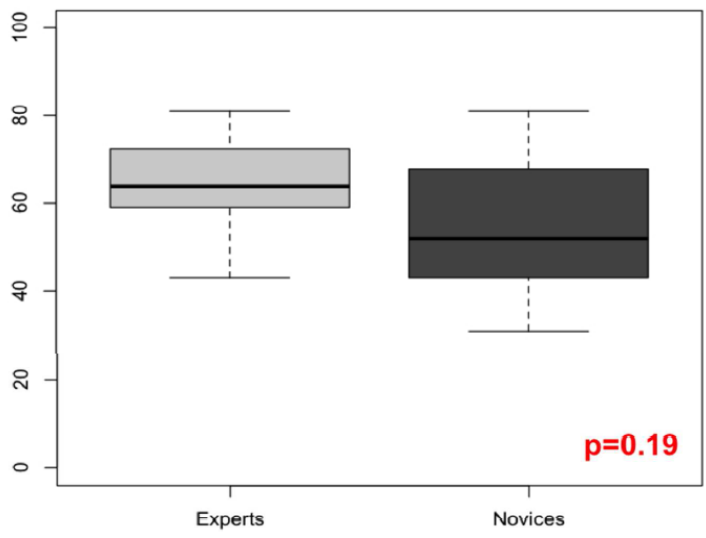